\documentclass[amssymb,amsmath,aip]{revtex4-2}

\usepackage[english]{babel}
\usepackage{graphicx}
\usepackage{color}
\usepackage{hyperref}
\usepackage{bm}
\usepackage{amsmath}
\usepackage{amssymb}

\DeclareGraphicsExtensions{.eps,.pdf,.png,.jpg}

\begin{document}
	
	
\title{Evolution of the electron distribution function during gas ionization by a sub-nanosecond microwave pulse of hundreds MW power. }
	
\author{Y. Bliokh, V. Maksimov, A. Haim, A. Kostinskiy, J. Leopold, and Ya. E. Krasik}
\affiliation{Physics Department, Technion, Israel Institute of Technology, Haifa 320003, Israel}
	
\begin{abstract}
	
The electron velocity distribution function in the plasma,
formed by gas ionization with a sub-nanosecond, hundreds of megawatts power level microwave pulse, is studied by a theoretical model and by numerical 3D simulations, the results of which agree well and show that the distribution varies along the pulse as a decreasing power-law function at the rear of the pulse.  Experiments performed in a waveguide filled with  helium gas confirm that energetic (from several keV to several tens of keV) electrons remain in plasma long after the pulse has crossed the experimental volume. These electrons continue the gas ionization over extended times up to tens of nanoseconds.

\end{abstract}		
	
\maketitle

\section{Introduction}

Gas discharge in a strong microwave field (more than tens of MW and tens of nanoseconds) was studied intensively in the late 1980s. It was shown by numerical simulations, theoretical and experimental studies that the plasma created during the discharge is far from an equilibrium. In particular, the distribution function of the plasma electrons differs strongly from being Maxwellian. 

Sub-nanosecond microwave pulses of hundreds of MWs power level, which are now available\cite{Eltchaninov-2004, Rostov-2016,Shafir-2017},
produce dense (up to $10^{11}-10^{12}\,{\rm cm}^{-3}$) plasma during their propagation through a low-pressure (several to several tens Torr) neutral gas. The plasma density, created during the short time (less than 1 ns), continues to increase over an extended period of time (several tens nanoseconds\cite{Cao-2020}), that the \textit{act of creation} can be considered  to be instantaneous. The energy left in the plasma after the microwave pulse’s passage, is concentrated mainly as kinetic energy of electrons. Thus, it is important to evaluate the electron velocity distribution function \textit{behind} the pulse, i.e., after the microwave field termination.

In this article the model, developed earlier in \cite{Glasov-1988, Glasov-1989, Brizhinev-1990, Glasov-1993, Soldatov-2001, Kuzelev-2001} for microwave fields of constant amplitude, is augmented to include time-dependent field amplitudes. It is shown that the velocity distribution function of electrons in the plasma left behind the microwave pulse differs considerably from the distribution function within the pulse duration. 3D numerical simulations confirm the results of the model and experiments confirm the existence of high energy electrons long after the pulse has left the experimental volume.

\section{Basic assumptions}

Let the following assumptions, similar to those used in \cite{Glasov-1988, Glasov-1989, Brizhinev-1990, Bychenkov-1992, Glasov-1993, Soldatov-2001, Kuzelev-2001}, to be valid.

\begin{enumerate}
	\item {The energy $w_\sim$ of the electron oscillating in the microwave field is much higher than the ionization potential $\varepsilon_i$ of the atoms:
	\begin{equation}\label{eq1}
		w_\sim\gg\varepsilon_i.
	\end{equation}}
	\item{The microwave frequency $\omega$ is much large than the  ionization  frequency $\nu_i$:
	\begin{equation}\label{eq2}
	\omega\gg\nu_i.
	\end{equation} }
	\item{The initial velocity $v(t_0)$ of an electron that appears due to an ionization event at time $t_0$, is small  compared to its oscillatory velocity $v_\sim$ and can be neglected,   
	\begin{equation}\label{eq3}	
		v(t_0)=0.
	\end{equation} }
	\item{The electron energy is so large, and the pulse duration $t_{\rm pulse}$  so small, that the change in electron energy due to ionization can be neglected, i.e.,
	\begin{equation}\label{eq4}
	\nu_it_{\rm pulse}\varepsilon_i\ll w_\sim.
	\end{equation}}
	\item{ The amplitude of the oscillating electric field varies adiabatically,
	\begin{equation}\label{eq5}
	t_{\rm pulse}\omega\gg 1.
	\end{equation}}
\end{enumerate}

\section {The Model}

The ionization of a neutral gas in the fields of the TM$_{01}$ mode of a high-power microwave pulse propagating in a circular waveguide (this choice corresponds to the experimental conditions\cite{Bliokh-2024}) is studied below.  The longitudinal, $E_z$, and radial, $E_r$, electric fields field components of this mode form an elliptically polarized field. The values of these components are equal approximately in the middle between the waveguide axis and the wall, and the wave electric field $\mathbf{E}$ is circularly polarized, $E=(E_r+iE_z)= E_0\exp(i\omega t)$. The total electric field $|E|$ reaches its peak value approximately at the same position. Thus, it seems reasonable to restrict our consideration to this region where the electric field is circularly polarized. This assumption is not obligatory but simplifies the computations below.

The electron motion in the electric field ${\mathbf E}(t)$, rotating in the $(x,y)$-plane, is described by the equation
\begin{equation}\label{eq6}
	\frac{d{\mathbf v}}{dt}=\frac{eE_0}{m}({\mathbf e}_x\cos\omega t+{\mathbf e}_y\sin\omega t),
\end{equation}
where $E_0$ is the field amplitude, ${\mathbf e}_x$ and  ${\mathbf e}_y$ are unit vectors directed along X and Y axes, respectively. Let an electron appear at  $t_0$ due to an ionization event. By virtue of assumption \#3, the initial velocity of this electron is equal to zero, so that the solution of Eq.~(\ref{eq6}) reads as:
\begin{eqnarray}\label{eq7}
	v_x(t)=v_\sim\sin\omega t-v_\sim\sin\omega t_0,\nonumber\\
	v_y(t)=-v_\sim\cos\omega t+v_\sim\cos\omega t_0,
\end{eqnarray}
where $v_\sim=eE_0/m\omega$. The first terms on the right-hand-side of Eqs.~(\ref{eq7}) describe the oscillatory motion in the rotating electric field, while the second terms describe regular, \textit{non-oscillating drift} motion. The drift velocity depends  only on the field amplitude and phase at the instant of time $t_0$ when this electron appears. In what follows, it is convenient to use the dimensionless velocities $u=v/v_i$ and $\tilde{u}=v_\sim/v_i$, normalized to the ionization threshold velocity $v_i=\sqrt{2\varepsilon_i/m}$.

Let the field amplitude be a time-dependent function, and $t_{\rm pulse}\ll\omega$  the characteristic time of its variations. From assumption \#4, one can assume that the oscillatory part of the solution in Eq.~(\ref{eq7}) follows the amplitude changes, while the drift motion remains unchanged: 
\begin{eqnarray}\label{eq8}
	u_x(t)=\tilde{u}(t)\sin\omega t-\tilde{u}(t_0)\sin\omega t_0,\nonumber\\
	u_y(t)=-\tilde{u}(t)\cos\omega t+\tilde{u}(t_0)\cos\omega t_0,
\end{eqnarray}
where $t\geq t_0$.

The "production" of new electrons as a result of electron-impact ionization of atoms or molecules does not affect the movement of the primary electrons (see assumption \#4 above). So, expression Eq.~(\ref{eq8}) is valid throughout the entire impulse, and this is correct for all electrons which appear as a result of ionization. Thus, the distribution function $f(u)$ of electrons which appear at time $t_0$,  can be presented as a superposition of the Dirac $\delta$-functions:
\begin{equation}\label{eq9}
	f(u;t_0,t)=v_i^{-1}\sum_{t_0\leq t}\delta\left[u-\sqrt{\tilde{u}^2(t)+\tilde{u}^2(t_0)-2\tilde{u}(t)\tilde{u}(t_0)\cos\omega(t-t_0)}\right],
\end{equation} 
were $u$ is the absolute value of the dimensionless velocity, $u=\sqrt{u_x^2+u_y^2}$.   
Assuming, that ionization events are homogeneously distributed in time over the field period (so-called equidistribution of electrons over the ``initial
phases'' \cite{Arutyunyan-1979,  Brizhinev-1990}), one can average the distribution function (\ref{eq9}) over the field period:
\begin{equation}\label{eq10}
	\langle f(u;t_0, t)\rangle=\sum_{t_0\leq t}\frac{1}{2\pi v_i}\int_{-\pi}^\pi d\varphi\delta\left[u-\sqrt{\tilde{u}^2(t)+\tilde{u}^2(t_0)-2\tilde{u}(t)\tilde{u}(t_0)\cos\varphi}\right], 
\end{equation} 
where $\varphi=\omega(t-t_0)$. Relation
\[\int dx\delta[F(x)]=\sum_{x_\ast,\; F(x_\ast)=0} 1/|F^\prime(x_\ast)|\]
allows integrating in Eq.~(\ref{eq10}):
\begin{equation}\label{eq11}
	\langle f(u;t_0, t)\rangle=\sum_{t_0\leq t}\frac{2u}{\pi v_i\sqrt{u^2-[\tilde{u}(t)-\tilde{u}(t_0)]^2}\sqrt{[\tilde{u}(t)+\tilde{u}(t_0)]^2-u^2}}. 
\end{equation}
When the field amplitude is constant, i.e., $\tilde{u}(t)=\tilde{u}(t_0)=\tilde{u}_{0}$, then the expression Eq.~(\ref{eq11}) for the distribution function coincides with the well known function \cite{Glasov-1993, Soldatov-2001}: 
\begin{equation}\label{eq11a}
\langle f\rangle=\frac{2}{\pi v_i}\frac{1}{\sqrt{4\tilde{u}_0^2-u^2}}.
\end{equation}

\section{Calculation algorithm} 

Let us divide the microwave pulse into small time intervals of duration $\Delta t\ll t_{\rm pulse}$ and neglect the field amplitude variations within the interval. Namely, let us replace the continuous function of time $u(t)$ by the stepwise function $u(t_n)$ where $n$ is the interval number. 

The density $\Delta n_1$ of electrons created during the first time interval $(t_1,t_1+\Delta t)$ due to gas ionization, is 
\begin{equation}\label{eq12}
	\Delta n_1=n_gv_i^2n_0\Delta t\int du u\sigma(u)\theta(u-1)f_0(u).
\end{equation}
Here $n_g$ is the neutral gas density, $n_0$ and $f_0(u)$ are the density and distribution function of the seed electrons, respectively,  $\sigma$ is the ionization cross section, and $\theta(x)$ is the Heaviside step function. The averaged distribution function of these new electrons is  $f(u; t_1,t_1)=2/\pi v_i\sqrt{4\tilde{u}(t_1)^2-u^2}$. Hereafter the  sign of the averaging $\langle\ldots\rangle$ will be omitted.  

The density $\Delta n_1$ of the electrons, created during the first time interval, remains constant, while their distribution function varies in time in accordance with Eq.~(\ref{eq11}) and takes the value
\begin{equation}\label{eq13}
	f(u;t_1,t_n)=\frac{2u}{\pi v_i\sqrt{u^2-[\tilde{u}(t_n)-\tilde{u}(t_1)]^2}\sqrt{[\tilde{u}(t_n)+\tilde{u}(t_1)]^2-u^2}} 
\end{equation}	 
at the $n^{\rm th}$ time interval. Thus, the density $\Delta n_m$ of electrons, created during the $m^{\rm th}$ interval, and their distribution function $f(u; t_m,t_n)$  at the $n^{\rm th}$ interval, $n>m$, can be calculated as follows:
\begin{equation}\label{eq14}
	\Delta n_m=n_gv_i^2\int du u\sigma(u)\left\{\sum_{k<m}\Delta n_kf(u; t_k, t_m)\right\}\Delta t. 
\end{equation}
\begin{equation}\label{eq15}
	f(u;t_m,t_n)\equiv f_{m,n}=\frac{2}{\pi v_i}\frac{u}{\sqrt{u^2-a^2_{m,n}}\sqrt{b^2_{m,n}-u^2}},
\end{equation}
where $a_{m,n}=|\tilde{u}(t_m)-\tilde{u}(t_n)|$ and $b_{m,n}=[\tilde{u}(t_m)+\tilde{u}(t_n)]$.
The expression in the curled brackets in Eq.~(\ref{eq14}) is the electron distribution function $f(u,t)$ at the instant $t=t_m$.  	

Let $N$ be the number of the time interval at the end of the pulse, where the field decreases to a level that required for a gas ionization, i.e., $u(t_N)\geq 0.5$ (the maximal electron velocity is twice as large as the oscillatory velocity $\tilde{u}$) and $\tilde{u}(t_{N+1})< 0.5$. The distribution function of electrons,  which appear at the $n^{\rm th}$ interval, takes at $t=t_N$ the value
\begin{equation}\label{eq16}
	f(u;t_n,t_N)=\frac{2u}{\pi v_i\sqrt{u^2-[\tilde{u}(t_n)-0.5]^2}\sqrt{[\tilde{u}(t_n)+0.5]^2-u^2}} 
\end{equation}	   
and the density of these electrons is $\Delta n_n$. Thus, this \textit{final} distribution function $f_{\rm final}(u)$ at the end of pulse can be written as follows:
\begin{equation}\label{eq17}
	f_{\rm final}(u)=\sum_{n\leq N}\Delta n_nf(u;t_n,t_N).
\end{equation}
Ionization continues over a long period (compared to the pulse duration) after the termination of the microwave pulse at the expense of the energy transferred to the electrons by the microwave field.

\section{Distribution functions in stationary and time-dependent amplitude microwave fields}

As mentioned in Sect. III, the distribution functions for  constant and time-dependent amplitude microwave fields are very different. In order to demonstrate this difference, let us consider the evolution of the distribution  function of a set of electrons placed in the field at a certain instant of time $t_0$, homogeneously distributed along the oscillations of the field, $t_0\in(0, 2\pi/\omega)$ (mentioned in Sect. III as an equi-distribution of electrons over the initial phase). The initial velocities of each electron is zero.

\begin{figure}[tbh]
	\centering \scalebox{0.3}{\includegraphics{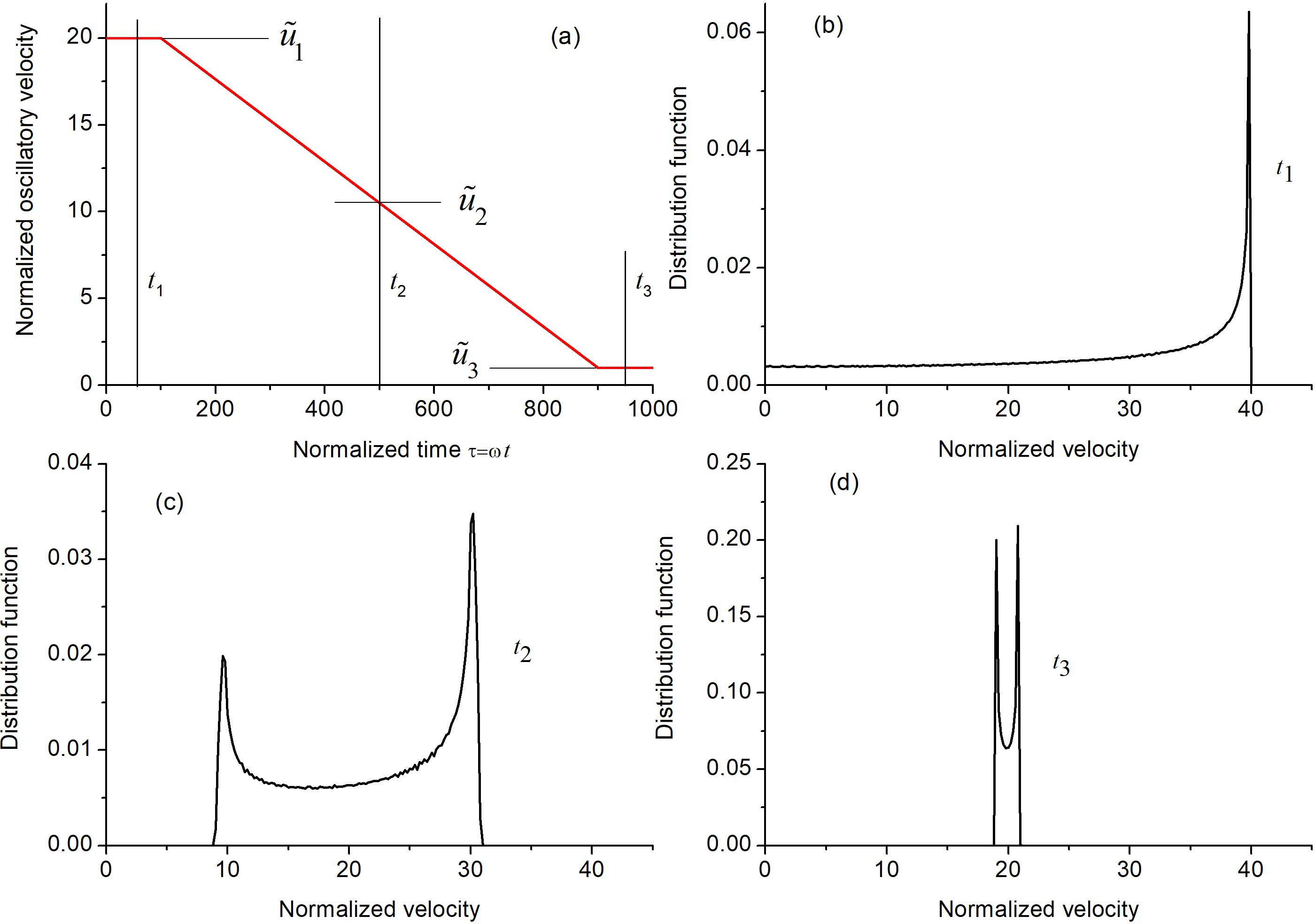}}
	\caption{(a) -- The oscillatory velocity $\tilde{u}(t)$ as a function of time. (b) -- The distribution function $f(u,t_1)$ in the field of constant amplitude (constant oscillatory velocity).  (c-d) -- The distribution functions at different instants of time $t_2$ and $t_3$. }
	\label{Particles}
\end{figure}

The oscillatory velocity $\tilde{u}$ (which is the measure of the field amplitude $E_0$) varies in time as shown in  Fig.~\ref{Particles}(a). Namely, the oscillatory velocity is constant during a certain time interval, $\tilde{u}(t)=\tilde{u}_1={\rm const}$, then, it decrease slowly to the value $\tilde{u}_3$, $\tilde{u}_1<\tilde{u}(t)<\tilde{u}_3$, after which it remains constant.    
Using the equation of motion Eq.~(\ref{eq8}), one can calculate numerically the 
velocity distribution function $f(u,t_i)$ of this set of electrons at any instant of time $t_i$. Results of these calculations are presented in Fig.~\ref{Particles}(b) and ~\ref{Particles}(c).

The time $t_1$ is chosen in the time interval where the oscillatory velocity is constant and equal to its value $\tilde{u}_1$ at the moment when the electrons are placed in the field. The distribution function $f(u,t_1)$ is shown in Fig.~\ref{Particles}(b). Its shape coincides with the shape obtained by expression Eq.~(\ref{eq11a}). The decrease of the oscillatory velocity is accompanied by a narrowing of the electron velocities interval from $0<u<2\tilde{u}_1$ at $t=t_1$ to $\tilde{u}_1-\tilde{u}_{2,3}<u<\tilde{u}_1+\tilde{u}_{2,3}$ at $t=t_{2,3}$ [see Fig.~\ref{Particles}(b)-(d)], in agreement with Eq.~(\ref{eq11}). As $\tilde{u}_3\rightarrow 0$, the distribution function collapses to a $\delta$-function, $f(u)\rightarrow\delta(u-\tilde{u}_1)$, as it follows from Eq.~(\ref{eq15}). {The reason of such behavior is very simple. Trajectory of the particle in $(u_x,u_y)$-plane, described by Eq.~(\ref{eq8}), is the circle of radius $\tilde{u}(t)$, whose center is located at the circle of radius $\tilde{u}(t_0)$. When  $\tilde{u}(t)\rightarrow 0$, then the trajectory shrinks to its center.  } 

This example demonstrates how strong is the difference between the electron distribution functions within the a microwave pulse with varying 
amplitude of the electric field and the constant amplitude case.

\section{Numerical results}

The parameters of the numerical simulations correspond approximately to the characteristics of the microwave pulses used in experiments\cite{Cao-2020}, namely, frequency 9.6 GHz, power 100-250 MW, and pulse duration of 0.5 ns at Full Width half Maximum (FWHM). The characteristic value of the oscillatory velocity is  $\sim5 \cdot 10^{9}-7\cdot10^9$ cm/s, that corresponds to $\sim (23-32)\,v_i$ for Hydrogen (ionization energy 13.6 eV, $v_i\simeq2.2\cdot 10^8{\rm cm/s}$) and  $\sim (17-24)\,v_i$ for Helium (ionization energy 24.6 eV, $v_i\simeq3\cdot 10^8{\rm cm/s}$).

For the ionization cross section $\sigma(v)$, let us use the Born approximation\cite{Landau-1977}:
\begin{equation}\label{eq18}
	\sigma(v)=\alpha\ln u/u^2.
\end{equation}
where $\alpha$ is the coefficient, which 
values for hydrogen, $\alpha_H$, and helium, $\alpha_{He}$, are\cite{Bell-1983} 
\begin{eqnarray}
	\alpha_H\simeq1.95\cdot 10^{-16}\,{\rm cm}^2,\nonumber\\
	\alpha_{He}\simeq1.88\cdot 10^{-16}\,{\rm cm}^2\label{eq19}
\end{eqnarray}

Using Eq.~(\ref{eq18}), equation (\ref{eq14}) can be written as follows:
\begin{equation}\label{eq21}
	\Delta n_m=n_g\frac{2v_i}{\pi}\alpha\Delta t\sum_{n<m}\Delta n_n\eta_{n,m},
\end{equation} 	
where 
\[\eta_{n,m}=\int_{a_{n,m}}^{b_{n,m}} du\frac{\ln(u)\theta(u-1)}{\sqrt{u^2-a^2_{n,m}}\sqrt{b^2_{n,m}-u^2}},\]
and $\theta(x)$ is the Heaviside step function\cite{GR}. It is convenient to present the neutral gas pressure $p$ in Torr and the time interval $\Delta t$ in nanoseconds. Then, Eq.~(\ref{eq21}) reads:   
\begin{eqnarray}
\Delta n_m\simeq p[{\rm Torr}]\Delta t[{\rm ns}]\sum_{n<m}\Delta n_n\eta_{n,m}\hspace{3mm}{\rm for\, H},\nonumber\\
\Delta n_m\simeq 1.3p[{\rm Torr}]\Delta t[{\rm ns}]\sum_{n<m}\Delta n_n\eta_{n,m}\hspace{3mm}{\rm for\, He}\label{eq22}
\end{eqnarray}
with $\Delta t=t_{\rm pulse}/N$.	
When all $\Delta n_n$ in the range $n\in(1,N)$ are calculated, one can determine the final distribution function $f_{\rm final}(u)$ of the electron velocities as  described by Eqs.~(\ref{eq16}) and (\ref{eq17}). Because the electron density is proportional to the seed electrons density $n_0$, it is sufficient to know the multiplication factor $\kappa=n/n_0$.

\begin{figure}[tbh]
	\centering \scalebox{0.3}{\includegraphics{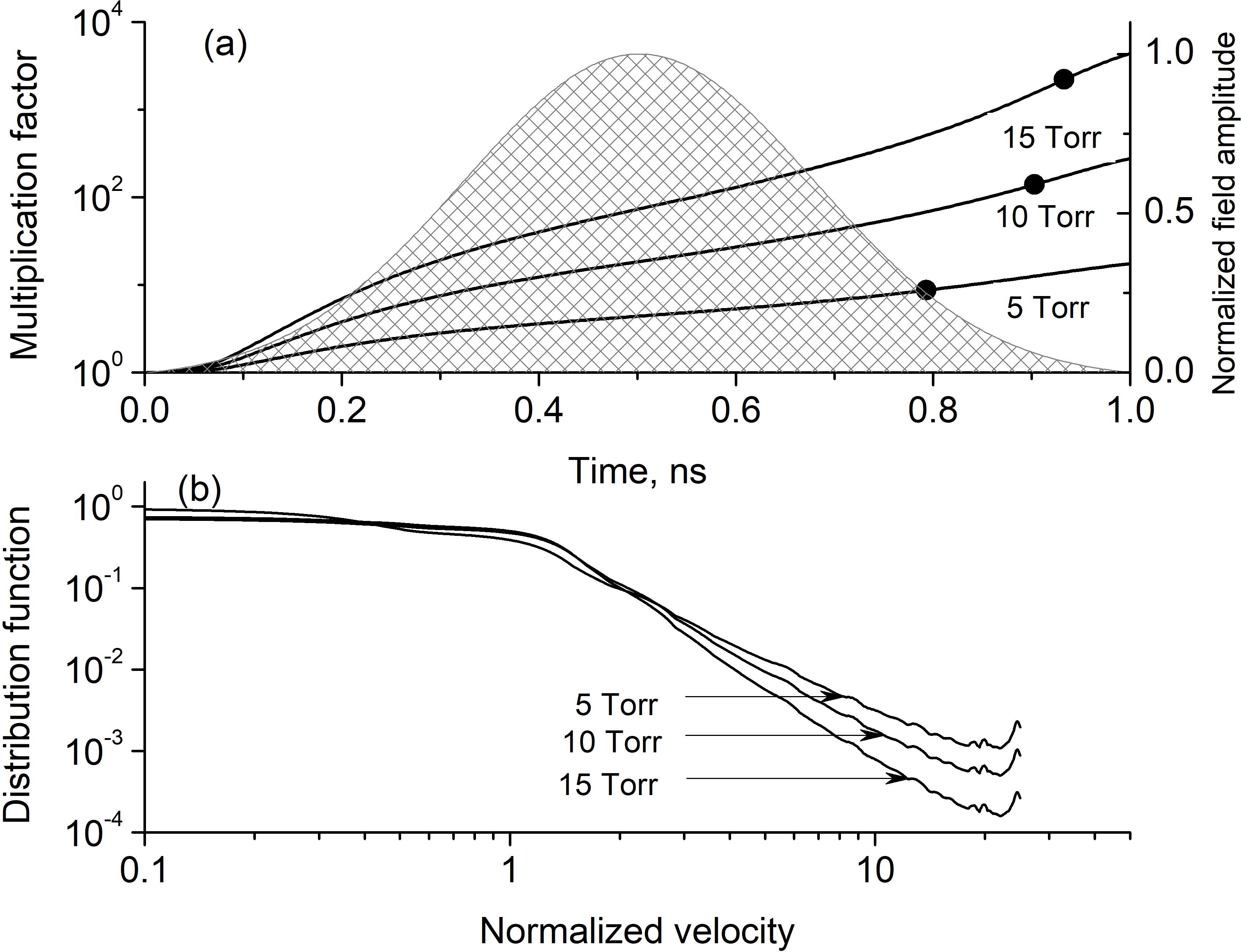}}
	\caption{Hydrogen ionization by a microwave pulse of 200 MW power (maximal oscillatory velocity $\tilde{u}=25$). (a) -- The electron multiplication factor $\kappa$ as a function of time. The normalized amplitude (left scale) of the microwave electric field is shown by the shaded region. Black points mark positions, where the density is half of its final value. Note, that half of the electrons are created near the rear part of the microwave pulse, where the field amplitude is small.  (b) -- Electron velocity distribution function at the end of the microwave pulse for different values of the gas pressure. }
	\label{FigA}
\end{figure}

As an example, the ionization of Hydrogen neutral gas (pressure 5-15 Torr) by a microwave pulse was studied numerically using the presented above algorithm. The pulse parameters were chosen to be: $\sim100-250$ MW power ($v_\sim/v_i\sim 20-30$), the pulse widths 0.5 ns at FWHM and 1 ns at full width at $v_\sim/v_i=0.5$ level. The results are shown in Fig.~\ref{FigA}.  

The distribution function, depicted in a log-log scale, is approximately constant at $v<v_i$, and decreases approximately linearly in the high-energy region $v>v_i$. It means that the distribution function decreases in this region as a power of the velocity, $f(v)\propto 1/v^\beta$, with the power $\beta$, which depends weakly on gas pressure. It is significant that the same behavior -- power-law dependence -- was observed in \cite{Shafir-2018}, where numerical study of a gas ionization by a powerful microwave pulse was carried out using 1D particle in cell (PIC) Monte Carlo collisional simulations, i.e., without assumptions mentioned in Sec. II.

The power-low decrease of the distribution function, which is much slower than  for a characteristic Maxwellian equilibrium distribution function, meaning that the microwave pulse leaves behind it's passage plasma with an abnormal amount of high-energy electrons. 

The fraction of energetic electrons increases with decreasing gas pressure.  This is because the main part of the electrons is created during the rear part of the pulse (see Fig.~\ref{FigA}a). The larger the gas pressure, the smaller the fraction of the pulse that contributes significantly to the electron density. Respectively, the smaller the field amplitude in this rear part of the pulse, the smaller  the energy of newly created electrons. 
{The electrons, created in the middle part of the pulse, where the oscillatory energy is maximal and remains large over an extended period, are responsible for the enhanced number of particles with velocity close to maximal.}

\section{3d numerical simulation}

The validity of the model described in the above sections was verified by 3D simulations  performed using the LSP (Large Scale Plasma) hybrid particle in cell (PIC) code \cite{LSP}. The simulation modeled a 250 MW, 9.6 MHz microwave pulse of $\sim0.5$ ns  FWHM propagating in a 1.4 cm radius waveguide in the TM$_{01}$ mode. The waveguide contained a 1 cm long slab filled with helium at 20 Torr pressure and a seed electron density of 10 cm$^{-3}$. The simulations account for only impact ionization according to known dependencies of the ionization cross section on electron energy and for elastic scattering cross section between electrons and atoms \cite{Cross-section}.  

\begin{figure}[tbh]
	\centering \scalebox{0.4}{\includegraphics{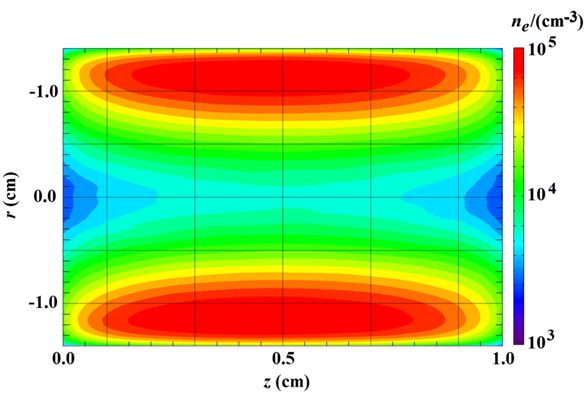}}
	\caption{The electron density distribution inside the gas filled slab at a moment after the microwave pulse has almost completely left the slab. }
	\label{e-density}
\end{figure}

In Fig.~\ref{e-density} the electron density distribution in the $(z,r)$ plane along the gas filled slab is drawn at a time when the microwave pulse has almost completely left the slab. One can see that the electron density is not uniform and has maximum density near the wall. The electric field decreases as it approaches the wall and consequently the oscillatory energy of the free electrons decreases to values corresponding to the largest ionization cross-section. 

The parameters and geometry of the 3D simulation correspond to the experimental conditions described in Section VIII. There are the number of distinctions between the simulation and the analytical model. The simulated electric field values and its polarization, and the electron density are very inhomogeneous across the waveguide cross section. The model assumes that the field is homogeneous and circularly polarized. Therefore, one can expect only a qualitative  correspondence between results obtained by the model and numerical simulation. 

\begin{figure}[tbh]
	\centering \scalebox{0.4}{\includegraphics{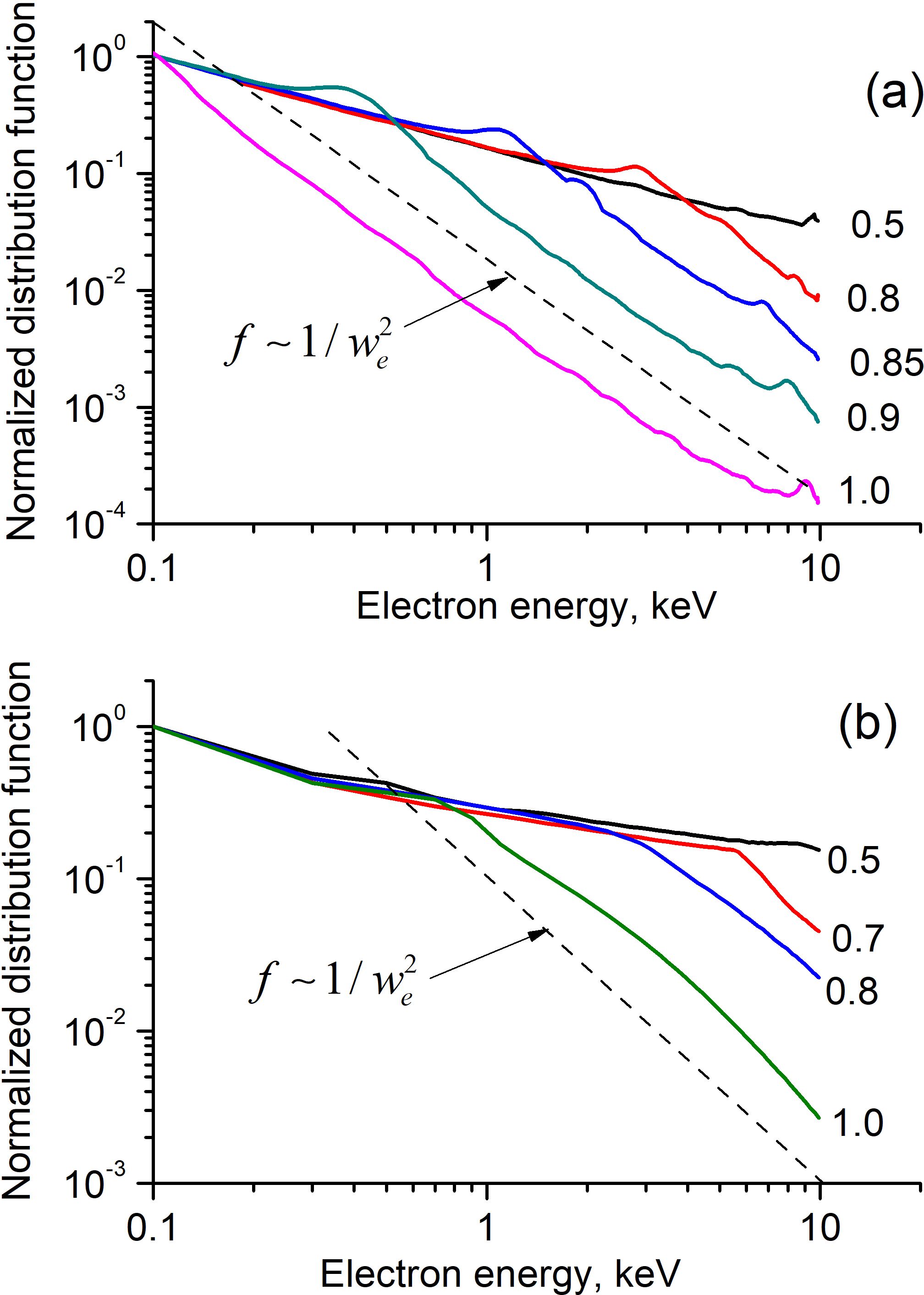}}
	\caption{The normalized electron energy distribution functions. (a) -- model, (b) -- 3d numerical simulation. The distribution functions are normalized in such a way that at electron energy 100 eV they are equal.  The curves are marked by the elapsed time $t$ from the beginning of the pulse. The time 0.5 ns corresponds to the pulse center, and $t=1$ ns corresponds to the pulse end. The dependence $f(w_e)\sim 1/w_e^2$ is shown by the dashed line. }
	\label{Comparison}
\end{figure}

One interesting feature of the distribution function evolution during the course of the microwave pulse follows from the model. The energy distribution function is a decreasing power-low function of the electron energy, $f(w_e, t)\sim 1/w_e^{\alpha(t)}$. The power $\alpha(t)$ increases sharply to the value $\alpha\simeq 2$ at a certain energy $w_c$, the value of which decreases with time. This evolution of the energy distribution function is shown in Fig.~\ref{Comparison}a. 

The results of the 3D simulation shown in Fig.~\ref{Comparison}b demonstrates the same specific evolution of the distribution function. Taking into account the differences between the initial conditions, assumed in the mathematical model and  the parameters of the numerical simulations, the comparison between the two is  satisfactory and confirms the dynamics implied by the model. 
 
\section{Experimental results}

It is very challenging to measure time-resolved energy spectrum of electrons generated during the ionization of a neutral gas by sub-nanosecond high-power microwave pulse. This is due to the time resolution limits of the detector and the necessity to use a window which separates the gas filled waveguide and the vacuum chamber where the detector can be placed. The finite thickness of the window restricts the minimal detectable energy of the electrons. Nevertheless, following the theoretical model and 3D simulations, experiments should confirm that the passage of a sub-ns high-power microwave pulse in a gas filled tube can create  seed electrons to produce observable ionization plasma and high energy electrons long after the pulse has left the experimental tube.

High energy electrons emission from the plasma, created by a microwave pulse in a waveguide filled with helium, was studied in the experimental setup described in earlier publications\cite{Cao-2023, Bliokh-2024, Cao-2024}. A backward wave oscillator operating in a super radiant mode produces the microwave pulse (TM$_{01}$ mode, up to 300 MW power, 9.6 GHz central frequency, and duration of 0.7 ns),  propagating in a 28-mm diameter circular waveguide contained in a larger diameter experimental chamber. Two calibrated loop-type couplers, installed near the entrance and exit of the waveguide register the incident, transmitted, and reflected microwave signals. The experimental chamber and waveguide were evacuated to a pressure of $10^{-3}$ Pa using turbo-molecular pump and then filled by helium to the desired pressure in the range 0.2 – 4 kPa. A 2-mm diameter optical fiber (Edmund optics)  with a scintillator (Saint-Gobain BC408, rise time of $\sim0.9$ ns and decay time of $\sim2$ ns), attached to its front end, is placed behind the Al foil at a distance of 10 mm from the 10-mm diameter hole made in the waveguide. At its other end, the fiber is attached to a fast ($\sim0.5$ ns rise time) photo-multiplier-tube (PMT) (Hamamatsu H10721–01). The fiber, scintillator and Al foils are placed inside a 6-mm-thick Al tube to protect the fiber from other possible sources of light and radiation. By varying the thickness of the Al foils (1–70 $\mu$m), the luminescence light, produced by electrons of sufficient energy to penetrate the given foil thickness, is registered. Using the collision stopping powers for electrons in aluminum\cite{Berger} the  energy of the electrons is determined.

\begin{figure}[h]
	\centering \scalebox{0.23}{\includegraphics{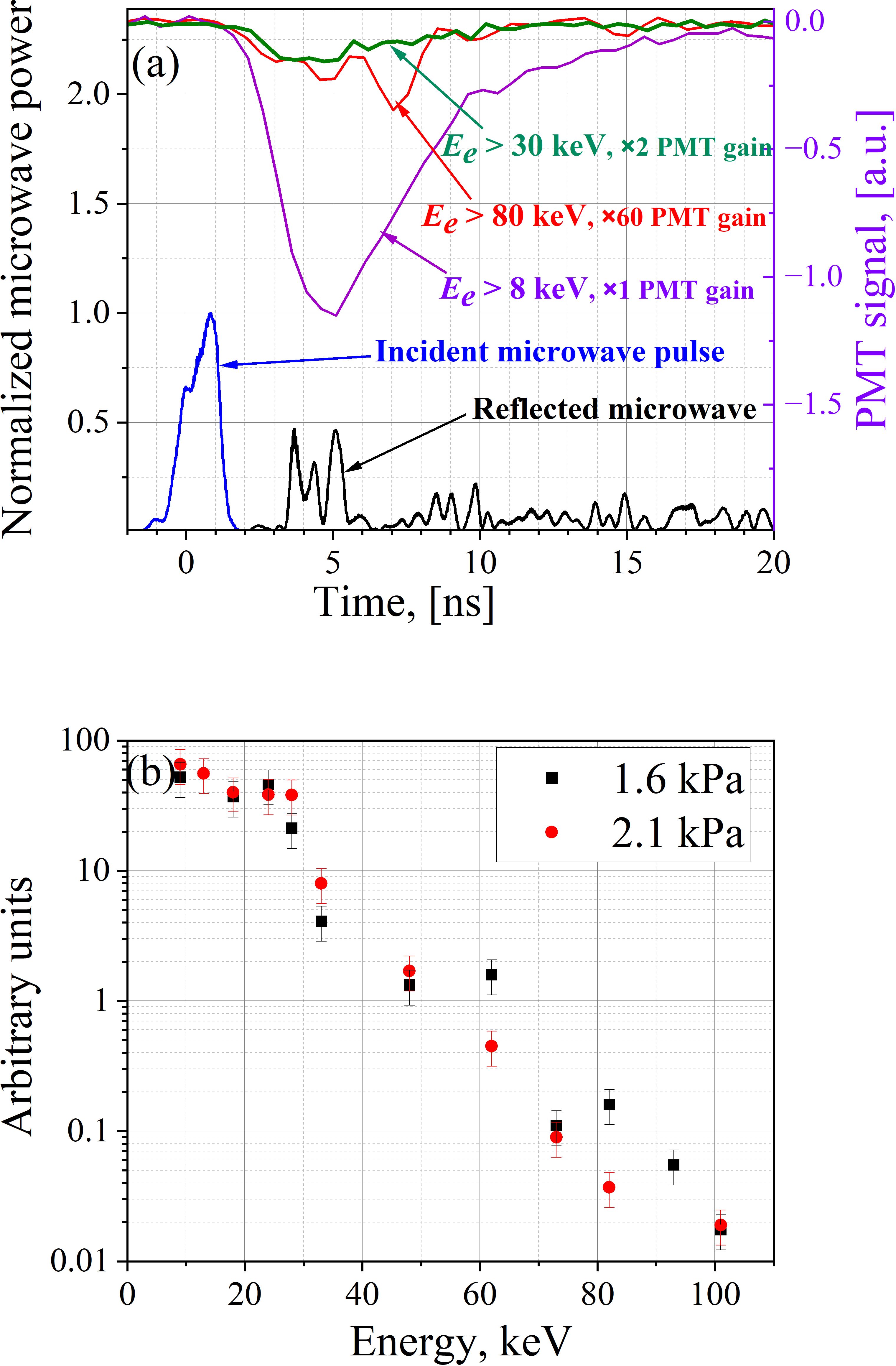}}
	\caption{(a) -- Typical waveforms of the plasma luminescence light intensity.  The waveguide is filled with Helium at 2 kPa pressure. The Al foils with 1-$\mu$m, 10-$\mu$m, and 50-$\mu$m thickness are placed in the front of the scintillator that corresponds to the electron energy $w_e>8$ keV, $>30$ keV and $>80$ keV, respectively. (b) Integral intensity of the luminescence light observed for different Al foil thickness in the range 1 $\mu$m – 70 $\mu$m versus the maximal stopping energy of electrons.}
	\label{Experiment}
\end{figure}

In Fig.~\ref{Experiment}a one can see typical waveforms of the luminescence light intensity obtained in experiments with 1.6 kPa He pressure for different thickness of Al foils, namely 1 $\mu$m, 10 $\mu$m and 50 $\mu$m which corresponds electron energies of $>$5 keV, $>$30 keV and $>$80 keV, respectively. Let us note that for 10 $\mu$m and 50 $\mu$m thick Al foils, the light emission is obtained when the gain of the PMT tube increases by 3 and 80 times, respectively, relative to the gain of the PMT for 1-$\mu$m thick Al foil. Thus, most of the energy spectrum of electrons appears in the energy range $<$10 keV. A small PMT signal was also registered with 80 $\mu$m thick Al foil, corresponding to  $\sim100$ keV electron energy. { In Fig. 5(b) The integral intensity of the luminescence light signals observed for different Al foil thickness in the range 1 $\mu$m – 70 $\mu$m versus the maximal stopping energy of electrons, is shown in Fig.~\ref{Experiment}(b). Qualitatively one can consider a fast decrease in the number of electrons with energy $>20$ keV. }

Thus, these experiments show that a powerful
sub-nanosecond microwave pulse produces plasma and energetic electrons which exists long after the pulse has left the experimental chamber.            

\section{Conclusions} 

A neutral gas ionization by a high-power, several hundreds of MW, sub-nanosecond microwave pulse can be considered as an instantaneous creation of  plasma in a strongly-nonequilibrium state. The energy, which remains in the plasma behind the pulse, is concentrated as the electrons energy. It has been shown in this paper, that the electrons accumulate such large energy, up to several keV, that gas ionization continues over a long period of time, up several tens of nanoseconds. The evolution of the plasma after the microwave pulse termination is determined by the electron velocity distribution function, which is modeled in this paper. {Particular attention has been given to the temporal evolution of the electron distribution function throughout the pulse. It has been shown that velocity distribution functions of electrons created under a neutral gas ionization by microwave field of constant or time-dependent amplitudes differs strongly. The electron distribution function behind the microwave pulse is a power-law function of the energy. This specific property has been obtained as in the developed model so as in the 3D numerical simulations.   }  

\begin{acknowledgements}
This study was supported by the PAZY Foundation, Grant No. 2032056.	
\end{acknowledgements}

\section*{Author declarations}
\subsection*{Conflict of Interest}
The authors have no conflicts to disclose.
\subsection*{Author Contribution}

{\bf Y. Bliokh:} Model (lead); Conceptualization (equal); Data curation (equal); Formal analysis (lead); Investigation (equal); Methodology (equal); Writing –original draft (lead); Writing – review \& editing (equal).
{\bf V. Maksimov:} Data curation (equal); Formal analysis (equal); Experimental Investigation (lead).
{\bf A. Haim:} Data curation (equal); Formal analysis; (equal); Investigation (equal).
{\bf A. Kostinskiy:} Data curation (supporting); Investigation (equal).
{\bf J. G. Leopold:} Software (lead); Validation (equal); Writing – review \& editing (equal).
{\bf Ya. E. Krasik:} Conceptualization (equal); Writing – review \& editing (equal); Funding acquisition (lead); Project administration (equal).

\section*{Data availability}

The data that support the findings of this study are available from the corresponding authors upon reasonable request.

\end{document}